\documentclass{optica-article}

\journal{opticajournal} 

\articletype{Research Article}

\usepackage{lineno}
\usepackage[output-decimal-marker={.}]{siunitx}
\DeclareSIUnit{\belmilliwatt}{Bm}
\DeclareSIUnit{\dbm}{\deci\belmilliwatt}
\DeclareSIUnit{\samplepersecond}{Sps}
\DeclareSIUnit{\decade}{dec}

\begin{document}

\title{High-precision automated setting of arbitrary magnitude and phase of Mach-Zehnder interferometers for scalable optical computing}

\author{Alessandro di Tria,\authormark{1,*} Gabriele Cavicchioli,\authormark{1} Pietro Giannoccaro,\authormark{1} Francesco Morichetti,\authormark{1} Andrea Melloni,\authormark{1} Giorgio Ferrari,\authormark{2} Marco Sampietro,\authormark{1} and Francesco Zanetto\authormark{1}}

\address{\authormark{1} Department of Electronics, Information and Bioengineering, Politecnico di Milano, piazza Leonardo da Vinci 32, Milano 20133, Italy\\
\authormark{2} Department of Physics, Politecnico di Milano, piazza Leonardo da Vinci 32, Milano 20133, Italy}

\email{\authormark{*}alessandro.ditria@polimi.it} 


\begin{abstract*} 
Photonic technologies offer promising solutions to the power consumption, bandwidth constraints and latency limits of electronic hardware used in high-performance computing and artificial intelligence. Recently, many studies have proposed and successfully demonstrated photonic accelerators based on integrated meshes of Mach-Zehnder interferometers (MZIs), enabling matrix-vector multiplications directly in the optical domain. While being fast and energy efficient, these photonic architectures still struggle to get the required precision for such applications, because setting the complex coefficients of MZI tunable gates with a high accuracy is still an unsolved problem. This work demonstrates high-precision automated setting and stabilization of MZI-based optical gates with a resolution of 7.01 and 8.04 bits for the output power and phase, respectively. Demonstration is achieved on a multistage silicon photonic circuit comprising a coherent input vector generator, an MZI matrix-vector multiplier, and a coherent receiver for phase measurement. The proposed control strategy can configure the MZIs to any desired working point, without any prior calibration or complex algorithm for the correction of hardware non-idealities, and prevents the propagation of programming errors, thus allowing scalability towards optical processors of large size. 

\end{abstract*}

\section{Introduction}\label{sec1}
High-performance computing based on deep learning and artificial intelligence (AI) \cite{lecun_2015_deep} algorithms is revolutionizing the field of data science, creating new solutions for a variety of problems, from sound \cite{ozer_2018_noise} and image \cite{krizhevsky_2017_imagenet} processing to biomedical engineering \cite{boulogeorgos_2020_machine}. Typically, these algorithms require intensive execution of matrix-vector or matrix-matrix multiplications (MVM and MMM, respectively), leading to high demands in terms of computational power, energy consumption and training time. Graphics processing units (GPUs), tensor processing units (TPUs), and application-specific integrated circuits (ASICs) appear to be the most appealing computing engines to manage this large amount of data, thanks to their intrinsic parallel processing capabilities \cite{wang_2019_benchmarking}. However, the bandwidth limitations ($<\SI{1}{\giga\hertz}$) and energy consumption per operation ($>\SI{0.4}{\pico\joule}$) \cite{cheng_2021_photonic} of these circuits limit further performance scaling at a reasonable cost, creating a significant bottleneck for the advancement of AI models and applications.

A promising platform for high-performance computing is offered by integrated photonics \cite{peserico_2023_integrated}. While a digital core shows, for a $N \times N$ matrix multiplication, a latency either proportional to $O(N)$ or approaching $O(\log{N}$) \cite{mallick_2020_rateless}, in a photonic core the calculation is performed within the transmission time of light in the chip, usually some tens of \si{\pico\second}. The overall time for the MVM should also take into account the electrical-to-optical (E/O) and optical-to-electrical (O/E) conversions (for the input generation and output sampling, respectively), which can be much faster than \SI{1}{\nano\second} and independent of the matrix size, making the photonic core latency $O(1)$. Another significant advantage of photonics for MVM is power efficiency. The transmission of light has inherently zero power consumption, unlike in electronics where the charging and discharging of electrical lines have a significant impact on the power budget, especially at high frequency \cite{miller_2017_attojoule}. The dominant source of power consumption in photonic cores lies in the weight setting and in the conversions from the electrical to the optical domain and vice-versa. The former scales as $O(N^2)$, however some photonic architectures, such as dynamic weight cross-bar arrays, and/or technological platforms (e.g. lithium niobate) allow for very low (\si{\femto\joule}) energy per weight \cite{youngblood_2022_coherent}. Therefore, the weight setting power consumption can be negligible with respect to the E/O/E conversions, which scale as $O(N)$, making a photonic core more efficient than its digital counterpart, especially as the size of the matrix scales up. Moreover, operating in the optical domain allows a further increase of the effective bandwidth by exploiting parallelism on multiple levels, using mode-, wavelength- and/or time-division multiplexing \cite{lu_2024_empowering,zhou_2023_memory,pappas_2025_reaching}.

Matrix-vector multiplication in the optical domain can be performed with free-space setups, using for example multi-plane light converters (MPLC) \cite{lin_2018_all}. However, these systems are bulky and their miniaturization can be technologically challenging. An integrated platform for MVM is offered by microring resonator (MRR) arrays \cite{ohno_2022_si}, where the input vector is encoded in multiple wavelengths and the weight banks are realized with tunable MRRs. This solution strongly reduces the footprint compared to the MPLC approach, but adds complexity in the management of multiple wavelengths, posing scalability challenges. Coherent architectures \cite{wu_2024_programmable} are instead designed to work with a single wavelength, making their operation easier. Among them, meshes of Mach-Zehnder interferometers (MZI) have been thoroughly studied and developed in recent years, since they can perform on-chip encoding and manipulation of complex numbers \cite{bandyopadhyay_2024_single,pai_2023_experimentally}. 

Although MZI meshes have proven to significantly accelerate MVM networks, an effort is still needed to realize reliable programmable photonic accelerators based on this approach. Indeed, setting the complex coefficients of the matrix implies precisely defining the working points of each interferometer, which is still an unsolved problem given the difficulty of measuring the local phase terms. In principle, a device-level pre-calibration stored in a look-up table could be employed \cite{bandyopadhyay_2021_hardware}. This approach can mitigate the effect of process errors, but it is scarcely reliable and accurate against thermal and other parasitic crosstalk effects, aging and generic environmental changes. All these effects limit the computation precision and the scaling of the processor order. 

To overcome this issue, in this paper we propose a novel control strategy to automatically set the optical power and phase difference at the outputs of an MZI and implement arbitrary complex-valued MVM with extreme accuracy and minimal insertion losses. The strategy is calibration-free, real-time and it can be seamlessly applied to large-scale MZI-based processors. The technique relies on a feedback control loop, which simultaneously sets the coefficients of the matrix and compensates for process tolerances and thermal fluctuations in real-time. We demonstrate the effectiveness of the strategy for a single MZI but, thanks to the independent setting of the parameters, the matrix size can be scaled up to realize high-order processors without correspondingly scaling the control complexity.

\begin{figure}[!b]
	\centering
	\includegraphics[width=0.5\textwidth]{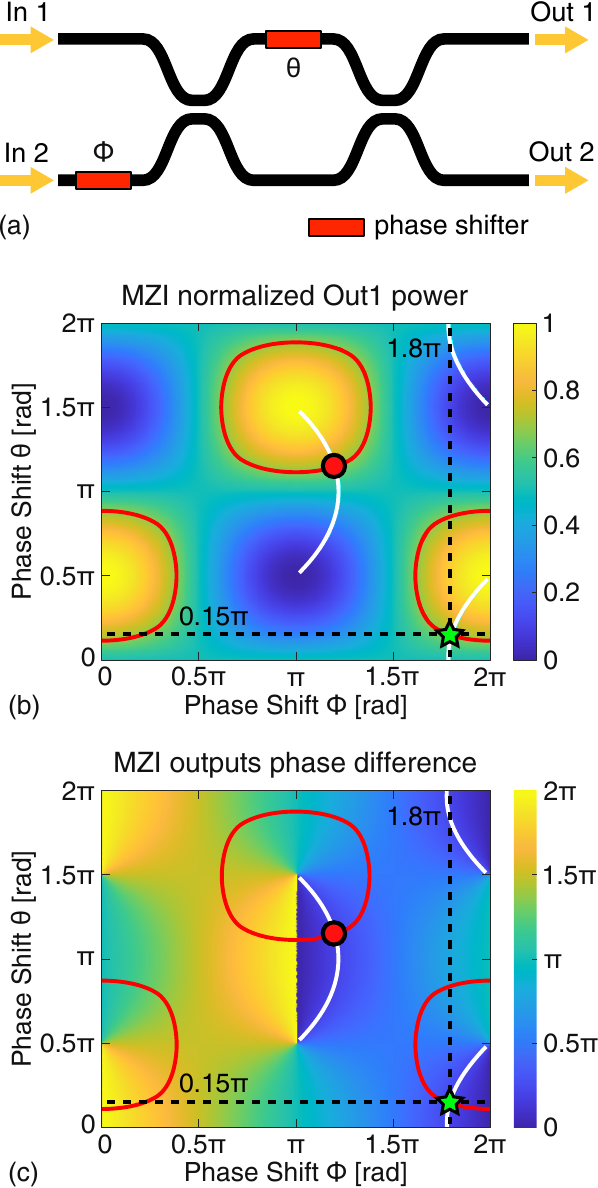}
	\caption{(a) Schematic of the MZI as a programmable 2x2 optical gate, with phase shifters $\phi$ and $\theta$. (b) Simulated power percentage and (c) phase difference at the output of the MZI with in-phase inputs of equal power, as a function of the phase shifts $\phi$ and $\theta$. The target working point, having $\phi = 1.8 \pi$ and $\theta = 0.15 \pi$, is highlighted by the green star. The red lines show the points sharing the same output power as the target, whereas the white lines highlight the points with the same phase difference. Two intersections are found, with the red dot indicating the wrong one.}
	\label{fig:fig1}
\end{figure}

\section{Control strategy}\label{sec2}
A Mach-Zehnder interferometer (Figure \ref{fig:fig1}a) with two 50/50 directional couplers can operate as a programmable $2\times2$ optical gate \cite{macho_2023_analog}, implementing a unitary matrix that can be configured using two phase shifters (PS) $\phi$ and $\theta$. Its transmission matrix $T_{MZI}$

\begin{equation}
        \label{eq:eq1}
            T_\text{MZI}=
            -je^{-j\theta/2}
            \begin{bmatrix}
                \sin(\frac{\theta}{2}) & \cos(\frac{\theta}{2})e^{-j\phi} \\
                \cos(\frac{\theta}{2}) & -\sin(\frac{\theta}{2})e^{-j\phi}
            \end{bmatrix}
\end{equation}
represents the unitary transformation of the elementary $2\times2$ gate. Higher-order unitary and non-unitary matrices \cite{cheng_2021_photonic} are implemented by combining multiple MZIs in a mesh topology \cite{reck_1994_experimental,clements_2016_optimal,mojaver_2023_addressing}. Accurately setting the complex-valued coefficients of such matrices corresponds to uniquely setting the phase shifts $\phi$ and $\theta$ induced by each actuator. This is a non-trivial operation, since the only available feedback signal is usually the light intensity information at the output of the entire processor. Therefore, the most common approach is to generate a calibration input vector, well-defined in power and phase, and monitor the processor output intensities as a function of each actuator state. In this way, a look-up table linking the PS values to the desired transfer matrix can be created. However, this calibration procedure is slow, especially when dealing with multiple MZIs in cascade, can be non-unique and requires periodic retraining to counteract the effect of thermal variations, aging or other long-term drifts. 

The opposite approach would be to independently monitor and set the status of each MZI output in intensity and phase with suitable on-chip sensors. However, it is interesting to notice that, even if this ideal condition applied, ambiguities would remain in setting the MZI transfer matrix. To understand this point, Figure \ref{fig:fig1}b and \ref{fig:fig1}c show the simulated normalized power and phase difference at the output of an MZI as a function of the PS $\phi$ and $\theta$, when the inputs are two optical beams of equal power and phase. Let us consider a target transfer matrix that can be obtained by setting $\phi = 1.8 \pi$ and $\theta = 0.15 \pi$. The green star in the figure shows the corresponding MZI configuration. If the interferometer output power is the only available information, ambiguity is found among all the possible working points indicated by the red lines, showing that the same power level can be achieved with different actuators commands. Measuring both power and phase difference at the output of the MZI would still result in an ambiguity between two solutions indicated by the green star and the red dot in Figure \ref{fig:fig1}b and \ref{fig:fig1}c. These solutions correspond to the intersections between the red and white lines, with the latter representing the working points that meet the target output phase difference constraint. Even though the two solutions share the same output power and phase difference, they implement two distinct transfer matrices $T_{MZI}$.

\subsection{Decoupling the effect of actuators}\label{subsec2.1} 
\begin{figure}[!t]
	\centering
	\includegraphics[width=0.5\textwidth]{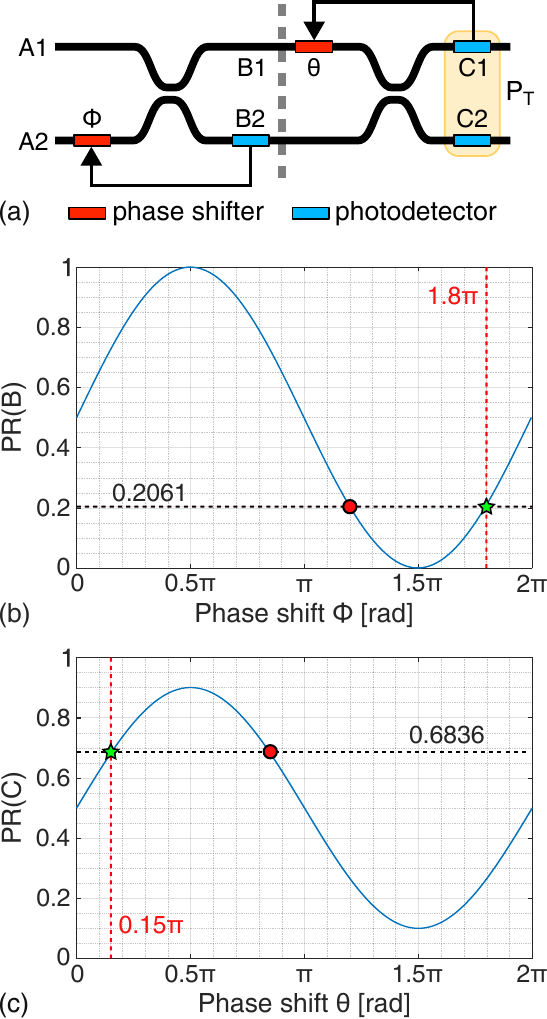}
 	\caption{(a) Schematic of the MZI conceptually divided into two cascaded PS-coupler-sensor sections each with its control loop. The total power $P_T$ is measured with the two output photodetectors. (b) Simulation of the power ratio in position $B_2$ as a function of the $\phi$ shift. A phase shift $\phi = 1.8\pi$ corresponds to a $PR(B) = 0.2061$ with positive derivative. (c) Simulation of the power ratio in $C_1$ as a function of $\theta$, when $\phi = 1.8\pi$. A phase shift $\theta = 0.15\pi$ corresponds to a $PR(C) = 0.6836$ with positive derivative.}
	\label{fig:fig2}
\end{figure}

One possibility to set the target working point precisely and without ambiguity is to detect the effect of each actuator separately. This can be achieved by considering the MZI as a cascade of two sections each composed of a PS, a directional coupler and a sensor to monitor the light intensity, with a local electronic control feedback loop. The schematic of this conceptual division is shown in Figure \ref{fig:fig2}a, where A, B and C indicate the positions of interest inside the MZI. To implement a calibration-free control strategy that is independent of the total optical power reaching the interferometer, the feedback loop can be designed to monitor and set the percentage of power (power ratio, PR) on the two outputs of each section. The power ratio in a generic section $X$ is defined as

\begin{equation}
   \label{eq:eq2}
     PR(X) = \frac{P(X_i)}{P_T}  
    \end{equation}
where $P(X_i)$ is the optical power detected by the PD in the corresponding position and $P_T$ is the total power traveling through the interferometer. Assuming negligible propagation and insertion losses within a single MZI, the total power $P_T$ can be measured in a single position, as shown in Figure \ref{fig:fig2}a where $P_T=P(C_1)+P(C_2)$ is assessed at the output of the device. In the rest of the paper, a normalized total power $P_T = 1$ is considered. For symmetry with respect to the actuator position, PR(B) is computed measuring the power in position B2, while PR(C) refers to the power in C1.

Assuming a generic input vector $V_A=[E_{A_1}, E_{A_2}e^{-j\chi_A}]^T$, where $E_{A_1}$ and $E_{A_2}$ represent the amplitudes of the optical fields and $\chi_A$ is their phase difference, the power ratio PR(B), measured by the PD at the output of the first section is related to the input vector $V_A$ as
    \begin{equation}
    \label{eq:eq3}
        PR(B)=\frac{1}{2}+ \frac{E_{A_1} E_{A_2}}{P_T} \sin(\chi_A+\phi).
    \end{equation}

Figure \ref{fig:fig2}b shows PR(B) as a function of the phase shift $\phi$, for an input vector $V_A=[1/\sqrt{2}, 1/\sqrt{2}]^T$. Considering the same example of the previous section, $\phi = 1.8\pi$ corresponds to PR(B)=0.2061 (green star). However, also $\phi =1.2\pi$ (red dot) generates the same PR(B). To discriminate the two solutions, the sign of the derivative of PR(B) with respect to $\phi$ can be exploited. The same considerations are valid for phase shifter $\theta$ and sensor $C_1$ (Figure \ref{fig:fig2}c), which measures PR(C). In the example, $\theta = 0.15\pi$ corresponds to PR(C) = 0.6836 with positive derivative. By operating two independent control loops to achieve the desired PRs and their derivatives, it is thus possible to automatically and independently set the phase shifts $\phi$ and $\theta$ without ambiguity and lock the MZI to the desired working point, implementing a unique unitary $2\times2$ matrix in a precise and stable way. 

\subsection{Transparent photodetectors}\label{subsec2.2}
\begin{figure}[!t]
	\centering
	\includegraphics[width=0.5\textwidth]{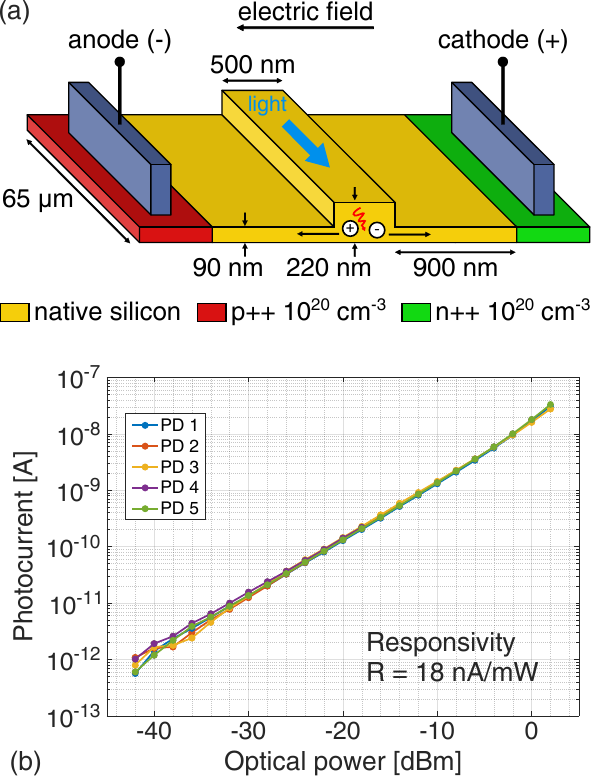}
	\caption{(a) Schematic of the transparent photodetector. For a length L=\SI{65}{\micro\meter}, the waveguide core is laterally extended and doped at its ends, to realize a transversal $p-i-n$ photodiode. (b) Measurement of the current of several photodetectors, as a function of the optical power. All curves are linear between \SI{0}{\dbm} and \SI{-40}{\dbm}, with a responsivity equal to \SI{18}{\nano\ampere/\milli\watt}.}
	\label{fig:fig3}
\end{figure}

In order to implement the proposed control strategy, light sensors need to be placed along each optical path, including the internal arms of the MZI. Absorbing photodetectors, like germanium photodiodes, are not suitable since they are characterized by a significant insertion loss. Minimally-invasive sensors are therefore needed. One possibility is to use detectors based on surface-state absorption (SSA), which is one of the natural causes of propagation losses in silicon waveguides \cite{baehr_2008_photodetection,grillot_2006_propagation}. Several examples of devices are found in the literature, including photodiodes \cite{chen_2009_cavity}, photoresistors \cite{perino_2022_high} and contactless integrated photonic probes (CLIPP) \cite{morichetti_2014_non,grimaldi_2022_non}. To avoid a complex calibration when computing the PR values and to make this calculation independent of the total on-chip optical power, the employed sensors should have a linear relation between light intensity and generated current. Photoresistors and CLIPPs generally show a sublinear response \cite{perino_2022_high,morichetti_2014_non}, therefore $p-i-n$ photodiodes have been chosen, enabling a calibration-free control strategy. 

Figure \ref{fig:fig3}a shows the schematic of the sensor and its dimensions. When light travels through the reverse-biased $p-i-n$ junction, the electron-hole pairs naturally generated by surface-state absorption are separated by the electric field in the device and collected at the contacts, producing a current that is proportional to the optical power in the waveguide. Since the waveguide core is not doped and the electrical contacts are placed sufficiently far away from it, the sensor does not introduce any additional insertion loss. It can thus be placed without penalties in any point of the photonic circuit where light monitoring is required, even in the internal arms of MZIs. Figure \ref{fig:fig3}b shows the measured responsivity curve of several transparent photodetectors in the same chip, highlighting a good linearity from \SI{0}{\dbm} down to \SI{-40}{\dbm} and a uniform responsivity of about \SI{18}{\nano\ampere/\milli\watt} for a \SI{65}{\micro\meter}-long sensor. The detectors dark current is around \SI{30}{\pico\ampere}, which is measured and subtracted from each readout to compute the correct PR value even in case of weak optical signals.

\section{Practical implementation}\label{sec3}
\begin{figure}[!t]
	\centering
	\includegraphics[width=\textwidth]{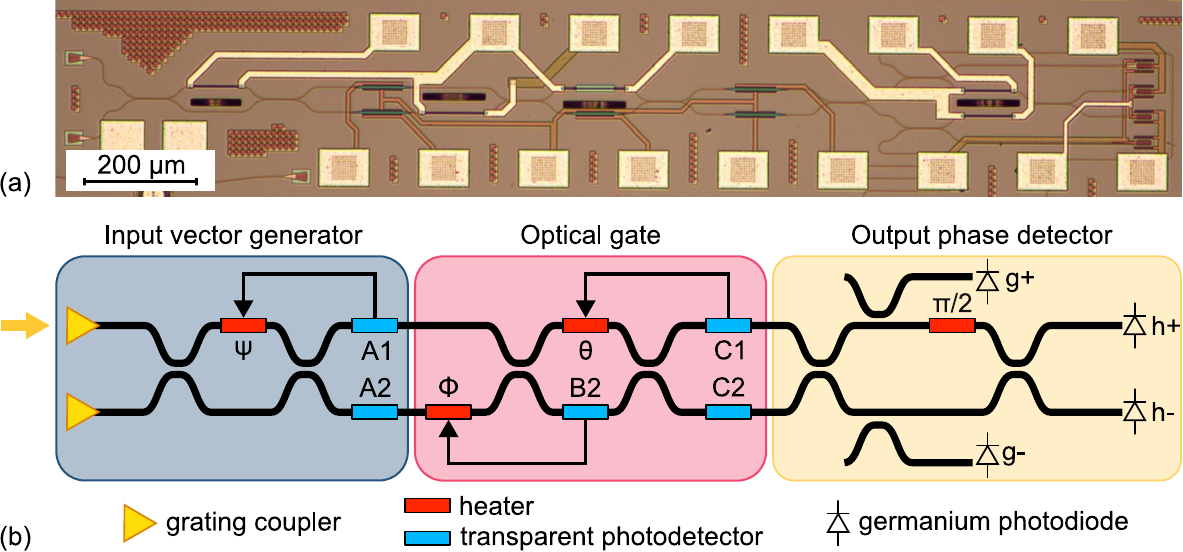}
	\caption{(a) Microscope photograph and (b) schematic view of the PIC. The first interferometer serves as input vector generator, the MZI in the middle is the optical gate with the proposed control strategy and the last stage is a coherent receiver, used to analyze the phase difference at the gate output. The local feedback loops used to control the three heaters are highlighted with black arrows.}
	\label{fig:fig4}
\end{figure}

The proposed control strategy has been tested on a specifically-realized silicon photonic integrated circuit. A microscope photograph and its schematic are shown in Figure \ref{fig:fig4}. The circuit includes an input vector generator, an optical processor and an output vector analyzer. Light at $\lambda = \SI{1550}{\nano\meter}$ is coupled from an external fiber to the chip through the top grating coupler (GC). A first balanced MZI is used to generate the input vector $V_A$ for the processor, by employing heater $\psi$ to control PR(A) and the relative phase between $E_{A_1}$ and $E_{A_2}$. For simplicity, this circuit topology can set only phase differences of 0 and $\pi$ but generalization to any phase is straightforward by adding another phase shifter. The generated vector serves as input for the optical gate, that implements the desired matrix. The optical processor features two heaters to set $\phi$ and $\theta$ and three transparent photodetectors in positions $B_2$, $C_1$ and $C_2$. A dummy photodetector is placed in the upper internal arm of the MZI (below heater $\theta$, not shown on the schematic) to keep the structure perfectly balanced. In order to validate the control strategy, the output optical vector $V_C=[E_{C_1}, E_{C_2}e^{-j\chi_C}]^T$ is measured in power by the photodetectors in positions $C_1$ and $C_2$, while the relative phase is assessed by a suitable phase analyzer, implemented with a coherent receiver. Indeed, the phase difference $\chi_C$ between the optical fields in positions $C_1$ and $C_2$ is equal to \cite{sun_2023_scalable}
\begin{equation}
    \label{eq:eq4}
    \chi_C=\tan^{-1} \biggl(\frac{g_+-g_-}{h_+-h_-}\biggl)
\end{equation}
where $g_+$, $g_-$, $h_+$ and $h_-$ are the intensities measured by PDs that, in our case, are on-chip integrated germanium photodiodes. A characterization of the phase detector is reported in Supplementary Section 1.

\subsection{Local feedback loops} \label{subsec3.1} 
\begin{figure}[!t]
	\centering
	\includegraphics[width=\textwidth]{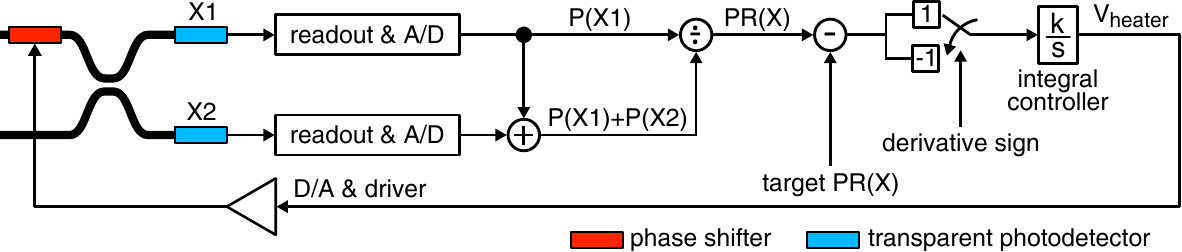}
	\caption{Schematic view of the local feedback used in both the optical gate and the input vector generator.}
	\label{fig:fig5}
\end{figure}

Each section, including PS, directional coupler and sensor, is controlled by the local feedback loop shown in Figure \ref{fig:fig5}, relating the power read by the transparent photodetector to the voltage used to drive the heater. Since each section features its own control loop, the complexity and latency of the feedback strategy are independent of the mesh size, providing a control solution that can be easily scaled to large photonic circuits. 

To implement the control algorithm, a custom electronic board (described in Supplementary Section 2) reads and digitizes the signal from the transparent photodetectors and feeds the information to a digital system. Here, the PR is computed according to Equation (\ref{eq:eq2}). The target PR is subtracted from the resulting value, and the difference is fed to a digital integral controller. The latter updates the heater voltage until its input, which is the loop error signal, is zeroed, thus locking the measured PR to the target value. The gain $k$ of the integral controller defines the bandwidth of the feedback loop, which is inversely proportional to its accuracy. In the following experiments, it has been set to obtain a control response time of \SI{25}{\milli\second}. 

As anticipated in Sec. \ref{subsec2.1}, the control strategy must be able to distinguish between points with opposite derivatives. The feedback loop automatically implements such distinction. Indeed, the sign of the derivative is assessed at run-time when computing the difference between the current PR and the target. Since the loop converges only when the overall feedback sign is negative, it is possible to select the solution with the desired derivative by inverting the error signal before the integral controller, as shown in Figure \ref{fig:fig5}.

\section{Experimental results}\label{sec4}

\subsection{Power control}
\begin{figure}[!t]
	\centering
	\includegraphics[width=\textwidth]{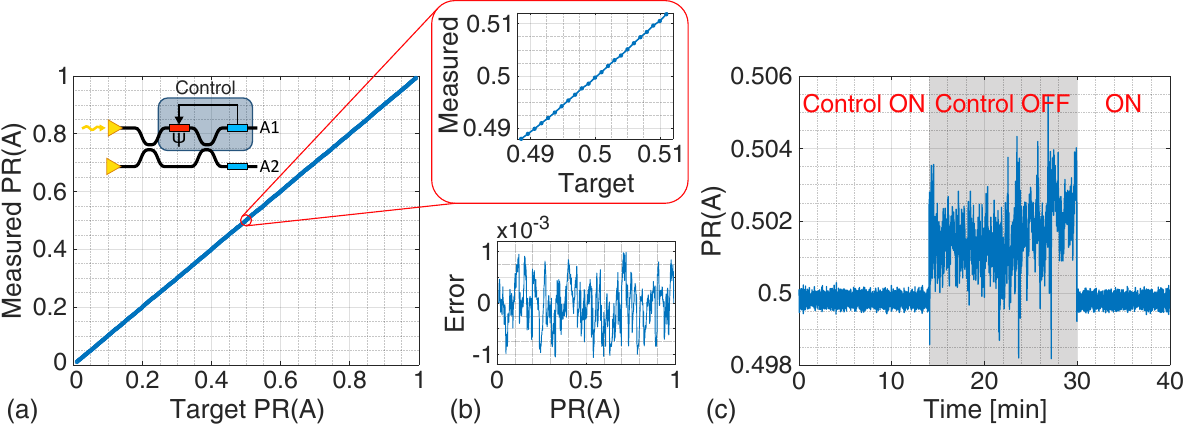}
	\caption{(a) Measured PR(A) vs target PR(A) over the entire range, with incremental steps of 0.1\%. A zoom of the measurement around PR(A) = 50\% is also reported. (b) Difference between measured and target PR(A), demonstrating a programming error always within 0.1\%. (c) Comparison of the measured PR over a long time span, when the control loop is on/off. First, a target PR = 50\% is set and actively kept for about \SI{17}{\minute}. Then, the control is switched off, causing a drift due to thermal variations and environmental factors. The system is then able to restore to the correct PR as soon as the control is switched on again.
    }
	\label{fig:fig6}
\end{figure}

The proposed control strategy has been tested starting with a single local feedback loop on the input vector generator of Figure \ref{fig:fig4}b, operating on the actuator $\psi$ and sensors in positions $A_1$ and $A_2$. PR(A) is computed using $A_1$ as the reference arm and the sum of the two photodetector signals to compute $P_T$. The target PR(A) has been swept from 0 to 1 with incremental steps of $0.1\%$, in order to validate the effectiveness of the control on the whole programming range. The measured PR(A) is shown in Figure \ref{fig:fig6}a. The feedback control is able to correctly set the target PR, showing excellent linearity and accuracy over the entire range with a root-mean-square error of only $0.046\%$, as highlighted in Figure \ref{fig:fig6}b.

Figure \ref{fig:fig6}c shows a measurement of the stability of the controlled working point over a period of 40 minutes. The control has been activated and kept for about 14 minutes to set PR(A) = 50\%. Afterward, the control has been switched off by keeping constant the driving voltage of heater $\psi$. For the successive 16 minutes, the measured PR shows higher oscillations and a significant drift, mainly due to thermal and environmental variations. The reactivation of the feedback loop after 30 minutes correctly restores the target PR, demonstrating the reliability of the implemented real-time control layer.

From these measurements, the metrics of accuracy and precision in setting a specific MZI output power can be extracted. In particular, accuracy $\sigma_{acc} = || PR_{meas} - PR_{target} ||$ is defined as the mean deviation from the target, while precision $\sigma_{prec} = \sqrt{\langle(PR_{meas} - PR_{mean})^2 \rangle}$ is computed as the standard deviation of repeated measurements from the mean value \cite{echeverria_2024_self}. Resolution is then defined as

\begin{equation}
    \label{eq:eq5}
    \text{Resolution} = \log_2 \left( \frac{PR_{max} - PR_{min}}{\sigma} \right).
\end{equation}

Extracting $\sigma_\text{acc}$ by averaging the error measurement in Figure \ref{fig:fig6}b and $\sigma_\text{prec}$ from the standard deviation in Figure \ref{fig:fig6}c, the obtained resolutions are 11.4 and 13.7 bits, respectively.

\subsection{Phase control}
The ability of the proposed control strategy to control the phase of the optical gate output has then been assessed. The gate requires two feedback loops, one using the information of PR(B) (see Figure \ref{fig:fig4}b) to drive heater $\phi$ and one relating PR(C) to heater $\theta$. An initial test has been run by activating only the control on heater $\theta$ to set PR(C)$ = 50\%$ and by driving heater $\phi$ manually from 0 to \SI{50}{\milli\watt}, corresponding to a phase shift of about $2\pi$ due to the rib-shaped waveguide below the heater. The optical gate has been stimulated with an input $V_A=[1/\sqrt{2},1/\sqrt{2}]^T$ set by the input vector generator.

\begin{figure*}[!t]
	\centering
	\includegraphics[width=\textwidth]{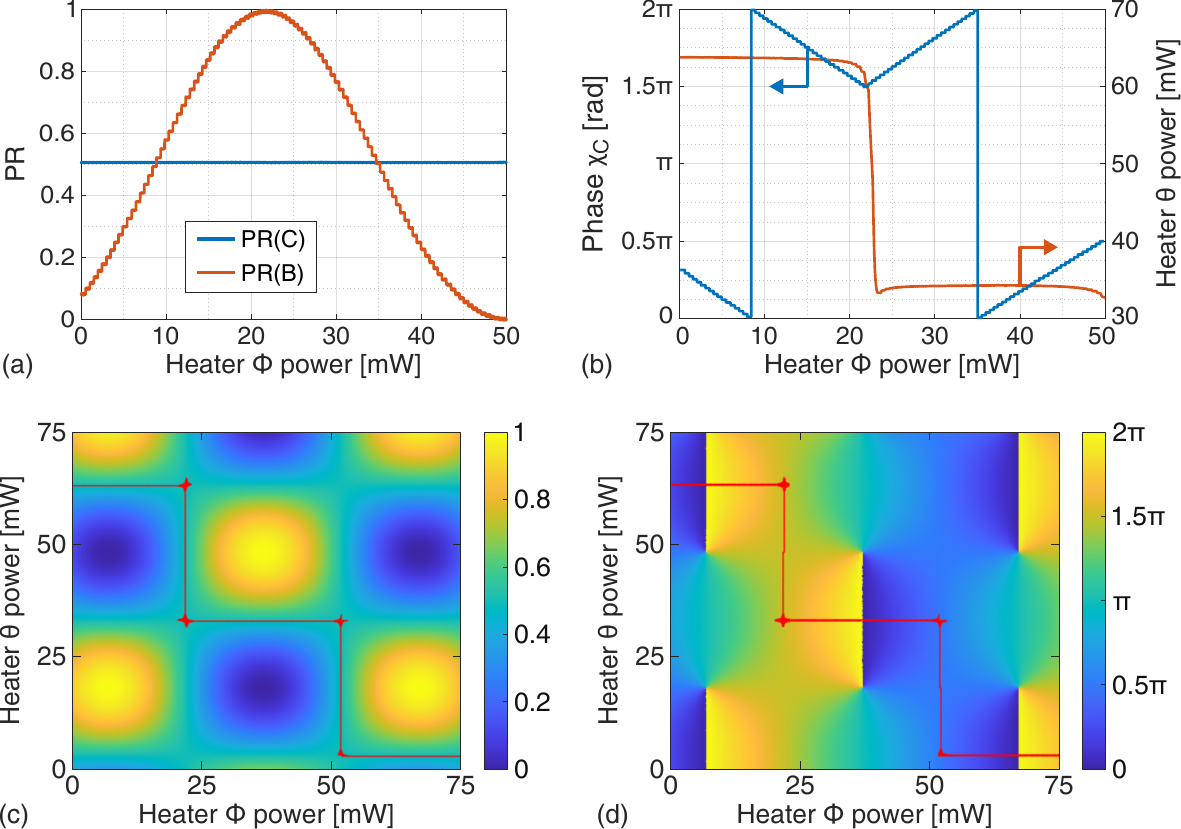}
	\caption{(a) Measured power ratios PR(B) (red) and PR(C) (blue) when the control of heater $\theta$ keeps PR(C) = 50\% and the heater $\phi$ power is swept from 0 to \SI{50}{\milli\watt}. (b) Measured output phase $\chi_C$ and command of the heater $\theta$ set by the control loop during the sweep. Simulation of the output (c) power and (d) phase of the MZI, as a function of the two heaters command. The red line indicates the points with PR(C)$= 50\%$, which match with the measurement shown in (b).}
	\label{fig:fig7}
\end{figure*}

Figure \ref{fig:fig7}a shows the optical power ratio measured internally (PR(B), red) and at the output (PR(C), blue) of the optical gate as a function of the heater power $\phi$, while the measured relative phase $\chi_C$ of the output vector $V_C$ is plotted in Figure \ref{fig:fig7}b. For the entire sweep of heater $\phi$, the output power ratio is kept at PR(C)$ = 50\%$ by the control loop while the output phase $\chi_C$  changes from about $0.5\pi$ to $1.5\pi$ and back, indicating that the MZI is changing its working point. To better understand this behavior, the evolution of heater $\theta$ is shown in Figure \ref{fig:fig7}b and the experimental measurement is superimposed to the simulated MZI transfer function in Figure \ref{fig:fig7}c and \ref{fig:fig7}d. For the first half of the sweep, the phase $\chi_C$ decreases until it reaches $1.5\pi$, while heater $\theta$ is kept at around \SI{64}{\milli\watt} by the control loop. Then, if the command of heater $\theta$ stayed constant, the derivative $\partial \text{PR(C)}/\partial\theta$ would change sign, as indicated by the simulation. However, since the control also preserves this information, the heater command suddenly changes to move to a working point with the same derivative sign. In this new operating region, the phase $\chi_C$ increases from $1.5\pi$ to $0.5\pi$, as evident both in the measurement and the simulation.

This result confirms that an additional control loop fed with the internal power ratio information is necessary to set the optical gate to a unique state, thus precisely defining an arbitrary target matrix.

\subsection{Matrix-vector multiplication}
The performance of two concurrent control loops has finally been measured. Once the input vector $V_A$ and the gate transfer matrix $T_{MZI}$ are set, the output of the optical gate corresponds to the product $V_C=T_{MZI} V_A$. Therefore, an input vector $V_A=[1/\sqrt{2},1/\sqrt{2}]^T$ of in-phase optical beams with equal power has been set using the input generator to configure the target matrix, by fixing PR(A) = 50\% and $\partial \text{PR(A)}/\partial\psi > 0 $. The target matrix has been chosen by selecting $\phi = 1.8 \pi$ and $\theta = 0.15 \pi$, as in Figure \ref{fig:fig1}, corresponding to PR(B) = 0.2061 and PR(C) = 0.6836. The two feedback loops of the optical gate, acting on heaters $\phi$ and $\theta$, have been activated to set these PR values with the correct derivatives.

According to Equation (\ref{eq:eq1}),
\begin{equation}
    \label{eq:eq6}
        T_{MZI}=
        \begin{bmatrix}
            -0.054 - 0.227j & 0.373 - 0.898j \\
            -0.229 - 0.945j & -0.089 + 0.216j 
        \end{bmatrix}               
\end{equation}
and hence $V_C= e^{-j 0.412\pi} [0.827  \ ,\   0.562 e^{-j 0.218\pi}   ]^T$ for $P_T=1$. In the following discussion, the common phase term $e^{-j 0.412\pi}$ is disregarded, since the output phase detector can measure only the phase difference between the two output beams.
 
After configuring the optical gate, the controls of $\phi$ and $\theta$ have been held to keep the matrix fixed and the control acting on $\psi$ has been used to generate different inputs and perform matrix-vector multiplications. Four different PR(A) have been selected, namely $20\%, 40\%, 60\%, 80\%$, as well as two phase difference values between the inputs, $\chi_A = 0$ and $\chi_A = \pi$. In total, nine different input vectors have been considered, including the calibration one. 

\begin{figure}[!t]
	\centering
	\includegraphics[width=\textwidth]{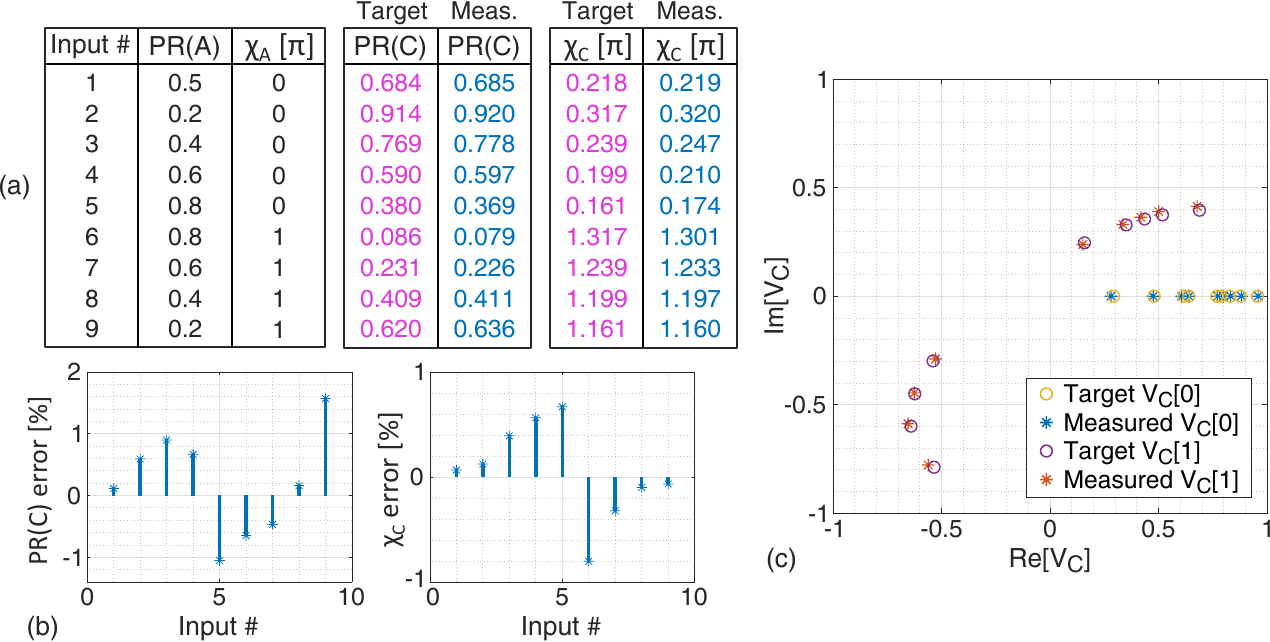}
	\caption{(a) Results of the MVM implemented with the proposed circuit for nine different input vectors, showing a comparison between the target (purple) and measured (blue) outputs. All vectors are expressed in terms of power ratio (PR) and phase difference ($\chi$). (b) Plots of the measurement error in power and phase, computed as the difference between the measured and target values. (c) Graphical representation of the two elements of the output vector $V_C$ for each MVM, plotted in the complex plane, compared to the theoretical results.}
	\label{fig:fig8}
\end{figure}

Figure \ref{fig:fig8} shows the measured results (blue) of the multiplication performed by the optical gate, compared to the theoretical values (purple). Panel (a) shows the input vectors and the target and measured output vectors, expressed in terms of PR(C) and phase difference $\chi_C$. The agreement is excellent for both amplitudes and phases, thus validating the control strategy. The difference between the target and measured values is reported in Figure \ref{fig:fig8}b, showing a root mean square error below 1\%. Using Equation (\ref{eq:eq5}), the resolution of the MVM output power and phase are equal to 7.01 and 8.04 bits (\SI{23.9}{\milli\radian}) respectively, better than similar architectures operated without feedback control \cite{kari_2024_realization}. From the measurements, the vector $V_C= P_T [\sqrt{PR(C)} \ , \ \sqrt{1-PR(C)} e^{-j \chi_C} ]$ is easily calculated and results are reported in the complex plane of Figure \ref{fig:fig8}c. Each marker represents a coefficient of the vector $V_C$ for the nine considered cases, very well matched with the exact results. 

It is worth noting that the output of the circuit are optical fields representing the result of the MVM, that can be used as inputs of another optical processor to implement further calculations. This result confirms that the processor works completely in the optical domain, without requiring any optical-electro-optical conversion that would degrade the power efficiency of the system. 

\section{Transparent phase measurement} \label{sec44}
\begin{figure}[!t]
	\centering
	\includegraphics[width=0.5\textwidth]{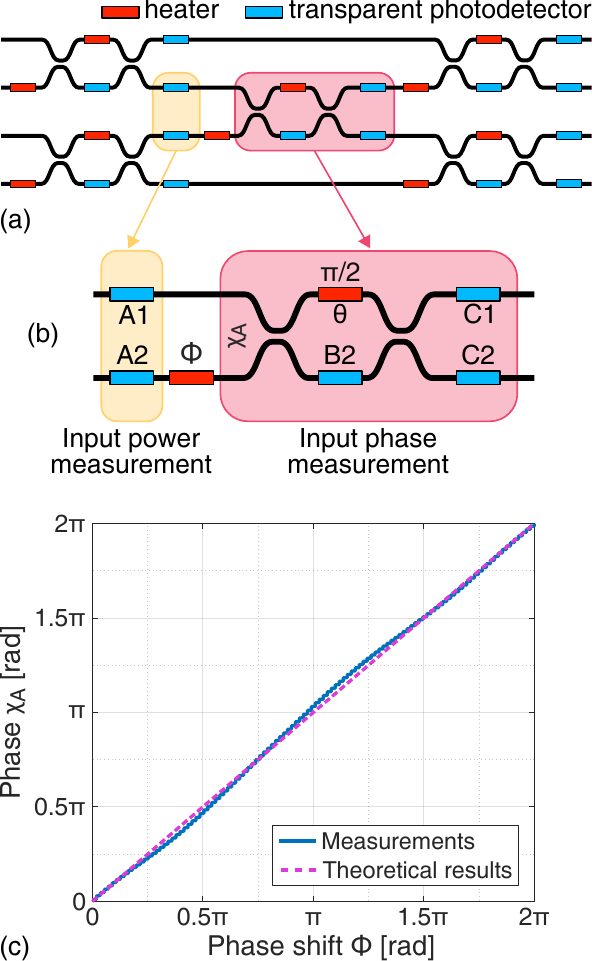}
	\caption{(a) Example of a 4x4 rectangular mesh in Clements topology \cite{clements_2016_optimal} employing the proposed control strategy. (b) Use of the interferometer as sensing element. The input power is measured with the output photodetectors from the previous MZIs, and the input phase difference $\chi_A$ is measured by setting $\theta = \pi/2$. (c) Measured $\chi_A$ with the optical gate in our PIC, as a function of the heater power $\phi$.}
	\label{fig:fig9}
\end{figure}

The proposed control strategy has been demonstrated on a first-order optical gate, to perform $2\times2$ matrix-vector multiplications with high accuracy. However, the approach can be straightforwardly extended to higher-order processors, as the $4\times4$ example shown in Figure \ref{fig:fig9}a. Seamless scaling is ensured by the independent MZI feedback loops based on transparent detectors. Indeed, it is enough to choose a well-defined mesh calibration input vector and compute the required power ratios after each directional coupler to easily configure the circuit. Although in principle this approach guarantees a correct configuration of the matrix, the possibility of assessing the state of each MZI in terms of both power and phase difference can still be useful to avoid the propagation of programming errors due to non-idealities in the circuit fabrication and layout. 

To this end, the transparent photodetectors placed at the output ports of each MZI stage can be used to measure the input optical power of the following stages, without requiring additional devices. The input phase difference can instead be assessed, before the gate configuration procedure, by using each MZI as a coherent phase detector. This is shown in Figure \ref{fig:fig9}b. By comparison with the coherent detector scheme shown in Figure \ref{fig:fig4}b, it is possible to derive

\begin{equation}
    \label{eq:eq7}
        \begin{cases}
            h_+ = P(C_1) \\
            h_- = P(C_2) \\
            g_+ = P_T - P(B_2) =P(C_1) + P(C_2) - P(B_2)\\
            g_- = P(B_2)\\
            \theta = \pi/2            
        \end{cases}               
\end{equation}
where it is assumed that the total power is preserved in each stage of the circuit ($P(B_1) + P(B_2)=P(C_1) + P(C_2)$) and that heater $\theta$ is set to generate a phase shift of $\pi/2$. The input phase difference $\chi_A$ is then computed as in Equation (\ref{eq:eq4})

\begin{equation}
    \label{eq:eq8}
    \chi_A =\tan^{-1}\biggl(\frac{P(C_1) + P(C_2) - 2P(B_2)}{P(C_1) - P(C_2)}\biggl)
\end{equation}

This idea has been experimentally validated by using the optical gate of the chip to measure the phase difference at its input. To this end, an input vector $V_A=[1/\sqrt{2},1/\sqrt{2}]^T$ has been generated with the first MZI and the heater power $\phi$ has been linearly swept to vary the phase from 0 to $2\pi$. The measured phase in Figure \ref{fig:fig9}b shows an excellent linearity, proving that the MZI gate can function both as a computing element and as a transparent phase sensor. This feature further increases the scalability of the proposed approach and enables independent configuration and verification of each MZI in any point of the processor.

\section{Conclusions and Outlook}\label{sec5} 
We demonstrated the precise and stable use of Mach-Zehnder interferometers as programmable computing elements. By controlling the actuators with suitable local feedback loops, it is possible to reliably configure any working point without any prior calibration or complex algorithm for the correction of hardware non-idealities. The control strategy has been validated, demonstrating very high accuracy (11.4 bits) and precision (13.7 bits) in setting a target optical power. The two concurrent local feedback loops have been successfully used to implement $2\times 2$ MVMs, as demonstrated by experimental measurements carried out with several input vectors. 

The proposed control strategy relies on the uniformity of the integrated photodetectors responsivities to correctly compute the PR values without requiring an individual calibration. In our design, we did not observe significant differences between devices, as demonstrated by the precision of the MVM results. It is worth noticing that local sensors uniformity is the only condition required to carry out an effective control action, even in larger architectures with multiple interferometers. Indeed, the information needed to configure each MZI is acquired with its three dedicated photodetectors. This feature significantly relaxes the requirements of the technological platform, since sensors uniformity is needed only at the local level of a single interferometer.

The presented approach can thus be seamlessly extended to high-order optical processors with no penalties on signal integrity and complexity, enabling several novel applications. With one feedback loop setting the internal phase shift $\theta = \pi/2$, the optical gate can be configured to measure its input phase difference, enabling its application as a direction-finding sensor \cite{zelaya_2024_photonic}, where the measured phase is directly related to the angle of arrival of a free-space optical beam. When both loops are activated, the processor can be used in any application where arbitrary unitary matrix transformations are needed, such as: neuromorphic photonics \cite{moralis_2022_neuromorphic}, where the MZI implements the linear stage of neural networks; quantum computing \cite{kwon_2024_quantum}, where the phase shifts express the probability of single photons to cross the quantum gate; optical routing and re-phasing, where an incoming beam sampled in multiple waveguides can be equalized in power and phase.

\section*{Acknowledgements}

The authors acknowledge Polifab, the nanofabrication facility of Politecnico di Milano (Italy), for assembling the samples. 

\section*{Disclosures}

The authors declare no conflicts of interest.


\bibliography{bibliography}

\newpage

\renewcommand{\thefigure}{S\arabic{figure}}
\renewcommand{\theHfigure}{Supplementary\thefigure}
\setcounter{figure}{0}
\setcounter{section}{0}

\section*{\centering\Large{SUPPLEMENTARY MATERIAL}}

\vspace{0.5cm}

\section{Coherent phase detector characterization}
\begin{figure}[!h]
	\centering
	\includegraphics[width=\textwidth]{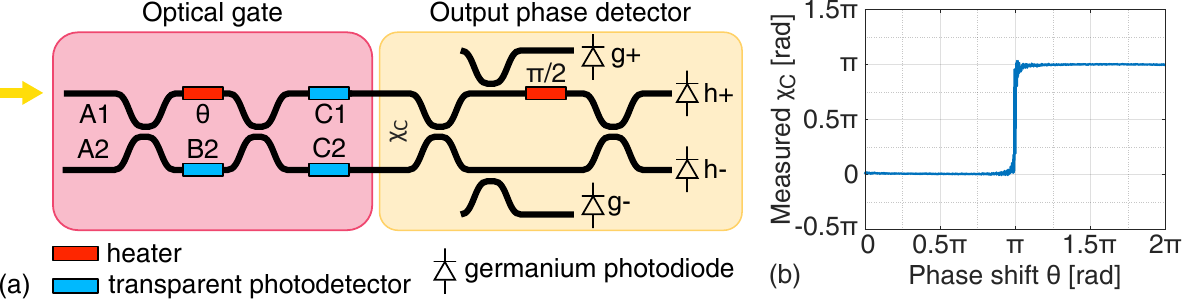}
	\caption{(a) Schematic of the optical gate and the output phase detector, with light injected only on the top waveguide in position $A_1$. (b) Measurement of the output phase difference $\chi_C$ after calibration, correctly showing only two levels.}
	\label{fig:figsup1}
\end{figure}

The output phase detector (Fig. \ref{fig:figsup1}a) is the only structure of the PIC operating without a feedback control loop. For this reason, it is sensible to non-idealities such as fabrication mismatches, thus needing an initial calibration in order to be operated. This calibration is performed by configuring the optical gate in a working point where its output phase $\chi_C$ is known. In particular, by using the input stage to generate the vector $V_A = [1, 0]^T$, the optical core output depends only on the phase shift $\theta$, and is equal to $V_C = [\sin{\left(\theta/2\right)}, \cos{\left(\theta/2\right)}]^T$, apart the common phase term. Therefore, the output phase difference $\chi_C$ can only be equal to
\begin{equation}
    \centering 
    \chi_C= 
        \begin{cases}
            0& \text{if } 0 < \theta < \pi\\
            \pi& \text{if } \pi < \theta < 2\pi .
        \end{cases}
\end{equation}

This information has been used to calibrate the phase shift measured by the coherent detector. Fig. \ref{fig:figsup1}b shows the experimental measurement of the MZI gate output phase difference after calibration, showing, as expected, a flat behavior with only two levels for the entire dynamic range. The measurement thus certifies that the phase detector can be reliably used to evaluate the result of the MVM carried out by the chip.

\section{Electronic control platform}
\begin{figure}[!h]
	\centering
	\includegraphics[width=0.9\textwidth]{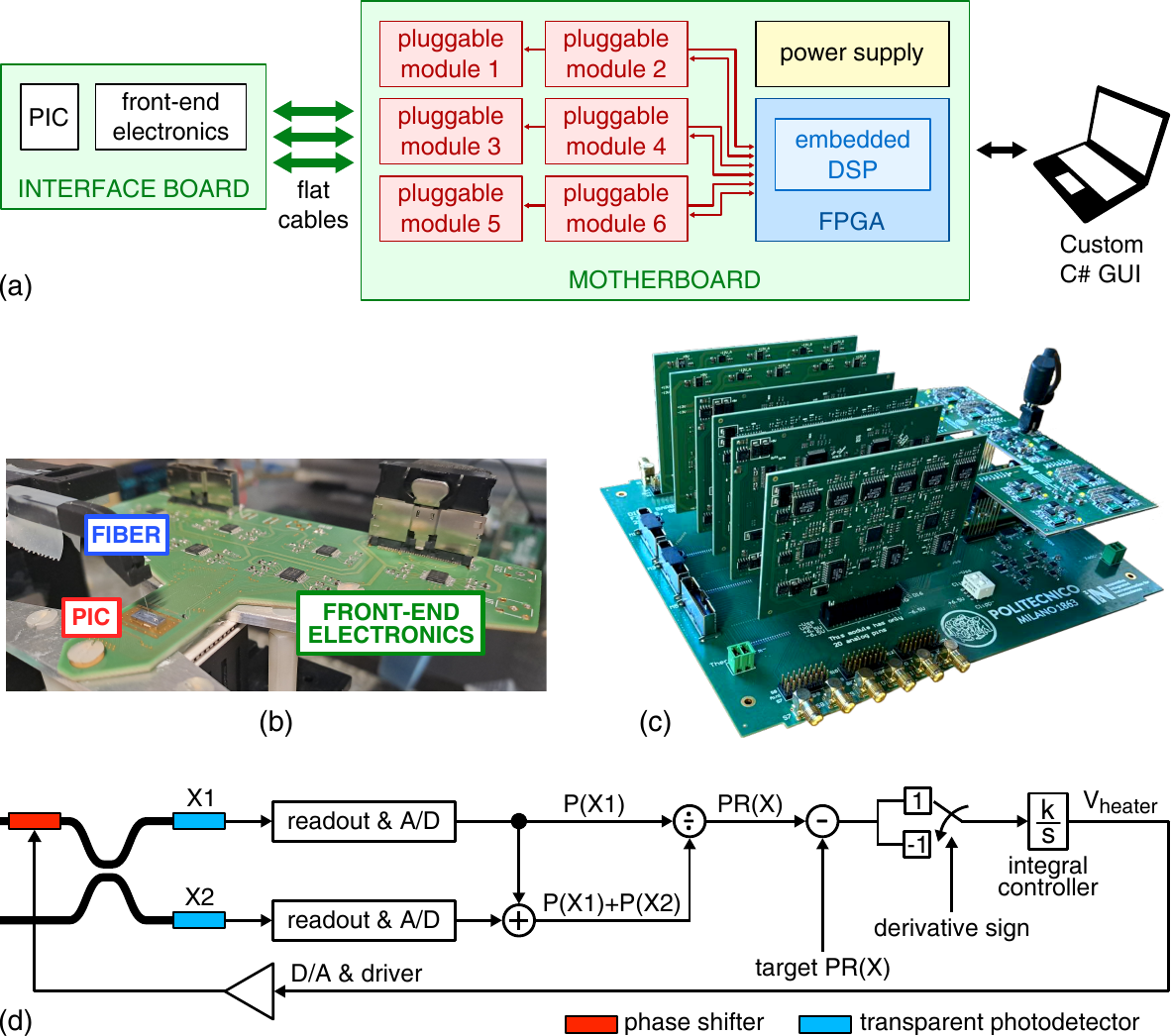}
	\caption{(a) Schematic view and (b), (c) photographs of the custom electronic system developed to implement the proposed control strategy. (d) Schematic of the control technique applied to a single actuator-coupler-sensor section. The implemented algorithm is repeated 3 times to control all MZIs of the PIC in parallel.}
	\label{fig:figsup2}
\end{figure}

In order to implement the proposed feedback strategy, a custom electronic platform has been developed. The system has a modular structure (Fig. \ref{fig:figsup2}a), comprising a small interface circuit board housing the PIC and a main control motherboard. 

The interface board (Fig. \ref{fig:figsup2}b) has been designed with an optimized shape to allow easy optical coupling and hosts the front-end circuits for amplification of the photodiodes current, placed close to the PIC to ensure a high-resolution measurement. The interface board is connected to the motherboard (Fig. \ref{fig:figsup2}c) with shielded cables. The latter hosts all the electronics needed for the control algorithm, with an FPGA core and up to six expansion modules dedicated to specific functionalities. In particular, the modules are designed to acquire the signals from the PIC and generate the voltages necessary for driving the actuators and biasing the sensors. The acquisition modules further condition the signals from the integrated photodiodes and digitize them with a set of 18-bit analog-to-digital converters. The FPGA then implements in the digital domain the feedback strategy, whose schematic is reported in Fig. \ref{fig:figsup2}d for convenience. The current power ratio (PR) is subtracted from the target and then accumulated, with proper sign, by the digital integrator. The bandwidth of the control loop can be tuned by adjusting the gain $k$ of the integral controller to achieve the desired update speed and accuracy. A bandwidth around \SI{100}{\hertz} is usually enough to compensate for thermal perturbations. The actuation modules finally bring the FPGA digital signals to the analog domain with a set of 16-bit digital-to-analog converters, followed by high-current drivers to supply the heaters. 

The whole system, with four acquisition and two actuation modules plugged into the motherboard, is able to read and condition up to 64 photodiodes and drive up to 48 heaters, and is therefore suitable to operate the entire PIC. The platform also allows further scaling to a higher-order processor, with up to 20 MZIs. To this end, the FPGA ensures maximum flexibility and reconfigurability, aided by a custom C\# graphical interface developed to configure the parameters of the system and visualize the state of the photonic processor on a personal computer.

\end{document}